\documentclass[aps,prl,twocolumn,groupedaddress,showpacs]{revtex4}

\begin{document}


\title{
Quantum Fermi's Golden Rule
}


\author{Fausto Rossi}
\email[]{Fausto.Rossi@PoliTo.It}
\affiliation{
Dipartimento di Fisica, Politecnico di Torino,
Corso Duca degli Abruzzi 24,
10129 Torino, Italy
}


\date{\today}

\begin{abstract}

We shall revisit the conventional adiabatic or Markov approximation, showing its intrinsic failure in describing the proper quantum-mechanical evolution of a generic subsystem interacting with its environment. In particular, we shall show that ---contrary to the semiclassical case--- the Markov limit does not preserve the positive-definite character of the corresponding density matrix, thus leading to highly non-physical results. To overcome this problem, we shall propose an alternative adiabatic procedure which (i) in the semiclassical limit reduces to the standard Fermi's golden rule, and (ii) describes a genuine Limblad evolution, thus providing a reliable/robust treatment of energy-dissipation and dephasing processes.

\end{abstract}


\maketitle



Present-day technology pushes device dimensions toward limits
where the traditional semiclassical or Boltzmann theory~\cite{ST} can
no longer be applied, and more rigorous quantum-kinetic
approaches are imperative \cite{QT}. However, in spite of the
quantum-mechanical nature of electron and photon dynamics in the core region
of typical solid-state nanodevices ---e.g., superlattices \cite{SL} and quantum-dot structures \cite{QD}--- the overall behavior of such quantum systems is often governed by a complex interplay between phase coherence and energy
relaxation/dephasing \cite{RMP}, the latter being also due to
the presence of spatial boundaries \cite{Frensley}. Therefore, a
proper treatment of such novel nanoscale devices requires a
theoretical modeling able to properly account for both coherent
and incoherent ---i.e., phase-breaking--- processes on the same
footing. 

The wide family of so-called solid-state quantum devices can be schematically divided into
two main classes: 
(i) a first one which comprises
low-dimensional nanostructures whose electro-optical response may be
safely treated within the semiclassical picture \cite{SP} (e.g., quantum-cascade lasers \cite{QCL}),
and (ii) a second one grouping solid-state devices
characterized by a genuine quantum-mechanical behavior of their
electronic subsystem (e.g., solid-state quantum logic gates \cite{QLG}) whose quantum evolution is only weakly disturbed by
decoherence processes.

For purely atomic and/or photonic quantum logic gates, decoherence phenomena are successfully described via adiabatic-decoupling procedures \cite{QO}
in terms of extremely simplified models via phenomenological parameters; within such  effective treatments, the main goal/requirement is to identify a suitable form of the Liouville superoperator, able to ensure/maintain the positive-definite character of the corresponding density-matrix operator \cite{QOS}. 
This is usually accomplished by identifying proper Lindblad-like decoherence superoperators \cite{QOS,Lindblad}, expressed in terms of a few crucial system-environment coupling parameters \cite{constrains}.

In contrast, solid-state devices are often characterized by a complex many-electron quantum evolution, resulting in a non-trivial interplay between coherent dynamics and
energy-relaxation/decoherence processes; it follows that for a
quantitative description of such coherence/dissipation
coupling the latter need to be treated via fully microscopic
models.

To this aim, motivated by the power and flexibility of the semiclassical kinetic theory \cite{ST} in describing a large variety of interaction mechanisms, a quantum generalization of the standard Boltzmann collision operator has been proposed \cite{RMP}; the latter, obtained via the conventional Markov limit, describes the evolution of the reduced density matrix in terms of in- and out-scattering superoperators.
However, contrary to the semiclassical case, such collision superoperator does not preserve the positive-definite character of the density-matrix operator.

To overcome this serious limitation, in this Letter we shall propose an alternative adiabatic procedure which (i) in the semiclassical limit reduces to the standard Fermi's golden rule \cite{FGR}, and (ii) describes a genuine Limblad evolution, thus providing a reliable/robust treatment of energy-dissipation and dephasing processes.

In order to discuss the main features and intrinsic limitations of the conventional adiabatic or Markov limit, let us recall its general derivation following the fully operatorial approach proposed in \cite{PRB}.
Given a generic physical quantity $A$ ---described by the operator
${\hat A}$--- its quantum plus statistical average value is given by
$A = {\rm tr}\left\{{\hat A} {\hat \rho}\right\}$,
where ${\hat \rho}$ is the so-called density-matrix operator. Its time evolution is dictated by the total (system plus environment) Hamiltonian. Within the usual interaction picture, the latter can be regarded as the sum of a noninteracting (system plus environment) contribution plus a system-environment coupling term:
${\hat H} = {\hat H}_\circ + {\hat H}'$; 
the corresponding equation of motion for the density-matrix operator ---also known as Liouville-von Neumann equation--- in the interaction picture is given by:
\begin{equation}\label{LvN_i}
{d{\hat \rho}^i \over dt} = -i \left[\hat{\cal H}^i, {\hat \rho}^i\right]\ ,
\end{equation}
where $\hat{\cal H}$ denotes the interaction Hamiltonian $\hat{H}'$ written in units of $\hbar$.

The key idea beyond any perturbation approach is that the effect
of the interaction Hamiltonian ${\hat H}'$ is ``small'' compared
to the free evolution dictated by the noninteracting Hamiltonian
${\hat H}_\circ$. 
Following this spirit, by formally integrating Eq.~(\ref{LvN_i}) from $-\infty$ to the current time $t$,
and inserting such formal solution 
for ${\hat \rho}^i(t)$ 
on the right-hand side of Eq.~(\ref{LvN_i}),
we obtain an integro-differential equation of the form:
\begin{widetext}
\begin{equation}\label{IDE}
{d \over dt} {\hat \rho}^i(t) = -i \left[\hat{\cal H}^i(t), {\hat \rho}^i(-\infty)\right]
-
\int_{-\infty}^t dt'
\left[\hat{\cal H}^i(t), \left[\hat{\cal H}^i(t'), {\hat \rho}^i(t')\right]\right]\ .
\end{equation}
\end{widetext}
We stress that so far no approximation has been introduced: Equations (\ref{LvN_i}) and
(\ref{IDE}) are fully equivalent, we have just isolated the first-order contribution from the full
time evolution in Eq.~(\ref{LvN_i}).

Let us now focus on the time
integral in Eq.~(\ref{IDE}). Here, the two quantities to be integrated over $t'$ are the interaction
Hamiltonian $\hat{\cal H}^i$ and the density-matrix operator ${\hat \rho}^i$. In the spirit of the perturbation
approach previously recalled, the time variation of ${\hat \rho}^i$ can be considered adiabatically slow
compared to that of the Hamiltonian $\hat{\cal H}$ written in the interaction picture, i.e.,
$\hat{\cal H}^i(t') = {\hat U}^\dagger_\circ(t') \hat{\cal H} {\hat U}^{ }_\circ(t')$;
indeed, the latter exhibits rapid oscillations due to the noninteracting evolution operator ${\hat U}_\circ(t) = e^{-{i{\hat H}_\circ t \over \hbar}}$.
As a result, the density-matrix operator ${\hat \rho}^i$ can be taken out of the time integral and
evaluated at the current time $t$.
Following such prescription, the second-order contribution to the system dynamics 
written in the Schr\"odinger picture 
for the case of a time-independent interaction Hamiltonian $\hat{\cal H}$ comes out to be:
\begin{equation}\label{LvN-eff}
{d{\hat \rho} \over dt} = 
-{1 \over 2} \left[\hat{\cal H},
\left[\hat{\cal K},{\hat \rho}\right]\right] 
\end{equation}
with
\begin{equation}\label{calK}
\hat{\cal K} 
= 
2 \int_{-\infty}^{0} dt' 
\hat{\cal H}^i(t') 
= 
2 \int_{-\infty}^{0} dt' {\hat U}^{ }_\circ(t') \hat{\cal
H} {\hat U}^\dagger_\circ(t') \ .
\end{equation}
As discussed extensively in \cite{PRB}, the operator $\hat{\cal K}$ describes 
energy-conserving scattering events (real processes) as well as energy-renormalization contributions (virtual processes); the latter are known to play a minor role, and in general may be safely neglected; such approximation amounts to impose the following  time-reversal symmetry on the system dynamics:
$\hat{\cal H}^i(t) = \hat{\cal H}^i(-t)$ \cite{renormalizations}.
Within such approximation scheme, the operator $\hat{\cal K}$ in (\ref{calK}) may be rewritten extending the time integration from $-\infty$ to $+\infty$:
\begin{equation}\label{calK-bis}
\hat{\cal K} = \int_{-\infty}^{+\infty} dt' {\hat U}^{ }_\circ(t') \hat{\cal
H} {\hat U}^\dagger_\circ(t') \ .
\end{equation}

The effective equation in (\ref{LvN-eff}) has still the double-commutator structure in (\ref{IDE}) but it is now local in time.
The Markov limit recalled so far leads to significant modifications
to the system dynamics: while the exact quantum-mechanical
evolution in (\ref{LvN_i}) corresponds to a fully reversible and
isoentropic unitary transformation, the instantaneous
double-commutator structure in (\ref{LvN-eff}) describes, in
general, a non-reversible (i.e., non unitary) dynamics characterized by energy dissipation and dephasing.
However, since any effective Liouville superoperator should describe correctly the time evolution of
$\hat\rho$
and since the latter, by definition, needs to be trace-invariant and positive-definite at any time,
it is imperative to determine if the Markov superoperator 
in (\ref{LvN-eff}) fulfills this two basic requirements.
As far as the first issue is concerned, in view of its commutator structure, it is easy to show that this effective superoperator is indeed trace-preserving.
In contrast, as discussed extensively in \cite{PRB}, the latter does
not ensure that for any initial condition the density-matrix
operator will be positive-definite at any time.
This is by far the most severe limitation of the conventional Markov approximation.

By denoting with $\{\vert \lambda \rangle\}$ the eigenstates of the noninteracting Hamiltonian $\hat{H}_\circ$, the effective equation (\ref{LvN-eff}) written in this basis is of the form:
\begin{equation}\label{LvN-eff-lambda}
{d\rho_{\lambda_1\lambda_2} \over dt} =
{1 \over 2} \sum_{\lambda'_1\lambda'_2}
\left[{\cal P}_{\lambda_1\lambda_2,\lambda'_1\lambda'_2}
\rho_{\lambda'_1\lambda'_2} 
-
{\cal P}_{\lambda_1\lambda'_2,\lambda'_1\lambda'_1} 
\rho_{\lambda'_2\lambda_2} \right] + {\rm H.c.} 
\end{equation}
with generalized scattering rates given by:
\begin{equation}\label{calP}
{\cal P}_{\lambda_1\lambda_2,\lambda'_1\lambda'_2} = {2\pi \over \hbar} H'_{\lambda_1\lambda'_1} H^{\prime *}_{\lambda_2\lambda'_2} \delta(\epsilon_{\lambda_2} - \epsilon_{\lambda'_2}) \ ,
\end{equation}
$\epsilon_\lambda$ denoting the energy corresponding to the noninteracting state $\vert \lambda \rangle$. 

The well-known semiclassical or Boltzmann theory~\cite{ST} can be easily derived from the quantum-transport
formulation presented so far, by introducing the so-called
diagonal or semiclassical approximation. The latter corresponds to
neglecting all non-diagonal density-matrix elements (and therefore
any quantum-mechanical phase coherence between the generic states
$\lambda_1$ and $\lambda_2$), i.e.,
$\rho_{\lambda_1\lambda_2} = f_{\lambda_1} \delta_{\lambda_1\lambda_2}$,
 where the diagonal elements $f_\lambda$ describe the semiclassical
distribution function over our noninteracting basis states.
Within such approximation scheme, the quantum-transport equation (\ref{LvN-eff-lambda}) reduces to the well-known Boltzmann equation:
\begin{equation}\label{BTE}
{d f_\lambda \over dt} =
\sum_{\lambda'} \left(
P_{\lambda\lambda'} f_{\lambda'} - P_{\lambda'\lambda} f_\lambda
\right)\ ,
\end{equation}
where
\begin{equation}\label{P}
P_{\lambda\lambda'} = {\cal P}_{\lambda\lambda,\lambda'\lambda'} =
{2\pi \over \hbar} |H'_{\lambda\lambda'}|^2 \delta\left(\epsilon_{\lambda}-\epsilon_{\lambda'}\right)
\end{equation}
are the conventional semiclassical scattering rates given by the well-known
Fermi's golden rule \cite{FGR}. 

At this point it is crucial to stress that, contrary to the non-diagonal density-matrix description
previously introduced, the Markov limit combined with the semiclassical or diagonal approximation ensures that at any time $t$ our semiclassical distribution function $f_\lambda$
is always positive-definite.
This explains the ``robustness'' of the Boltzmann transport equation (\ref{BTE}), and its extensive application in solid-state-device modeling as well as in many other areas, where quantum effects play a very minor role.
In contrast, in order to investigate genuine quantum-mechanical phenomena, the conventional Markov superoperator in (\ref{LvN-eff}) cannot be employed, since it does not preserve the positive-definite character of the density matrix $\rho_{\lambda_1\lambda_2}$.

As anticipated, aim of the present Letter is to propose an alternative formulation of the standard Markov limit, able to provide a Lindblad-like scattering superoperator, thus preserving the positive-definite character of our density matrix.
 To this end, let us go back to the integro-differential equation (\ref{IDE}). As previously discussed, the crucial step in the standard derivation is to replace $\hat\rho^i(t')$ with $\hat\rho^i(t)$. Indeed, since in the adiabatic limit the time variation of the density matrix within the interaction picture is negligible, the latter can be evaluated at any time. Based on this remark, what we propose is the following time symmetrization: given the two times $t'$ and $t$, we shall introduce the ``average'' or ``macroscopic'' time $T = {t+t' \over 2}$ and the ``relative'' time $\tau = t-t'$. The basic idea is that the relevant time characterizing/describing our effective system evolution
is the macroscopic time $T$. Following this spirit, it is easy to rewrite the second-order contribution in Eq.~(\ref{IDE}) in terms of the new time variables $T$ and $\tau$:
\begin{widetext}
\begin{equation}\label{IDE-new}
{d \over dT} {\hat \rho}^i(T) = 
-
\int_0^\infty d\tau
\left[\hat{\cal H}^i\left(T+ {1 \over 2} \tau\right), \left[\hat{\cal H}^i\left(T- {1 \over 2}\tau\right), {\hat \rho}^i\left(T-{1 \over 2}\tau\right)\right]\right]\ .
\end{equation}
\end{widetext}
Following again the spirit of the adiabatic decoupling, we shall now replace $\hat\rho^i\left(T-{1 \over 2}\tau\right)$ with $\hat\rho^i(T)$; the resulting effective equation rewritten in the original Schr\"odinger picture comes out to be:
\begin{equation}\label{LvN-eff-new1}
{d \over dT} {\hat \rho}(T) = 
-
\int_{0}^{\infty} d\tau
\left[\hat{\cal H}^i\left({1 \over 2} \tau\right), \left[\hat{\cal H}^i\left(-{1 \over 2}\tau\right), {\hat \rho}(T)\right]\right]\ .
\end{equation}
Neglecting again renormalization contributions (see note \cite{renormalizations}), 
Eq. (\ref{LvN-eff-new1}) may be rewritten by extending the time integration over $\tau$ from $-\infty$ to $+\infty$:
\begin{equation}\label{LvN-eff-new2}
{d \over dT} {\hat \rho}(T) = 
-{1 \over 2}
\int_{-\infty}^{+\infty} d\tau
\left[\hat{\cal H}^i\left({1 \over 2} \tau\right), \left[\hat{\cal H}^i\left(-{1 \over 2}\tau\right), {\hat \rho}(T)\right]\right]\ .
\end{equation}
 By Fourier expanding the above symmetric convolution integral we finally get the desired Lindblad-like scattering superoperator:
\begin{equation}\label{Lindblad}
{d {\hat \rho}\over dT} = 
-{1 \over 2}
\int d\omega \left[\hat{\cal L}(\omega), \left[\hat{\cal L}(\omega), {\hat \rho}\right]\right]
\end{equation}
with
\begin{equation}\label{calL}
\hat{\cal L}(\omega) = {1 \over \sqrt{2\pi}} \int_{-\infty}^\infty d\tau {\hat U}^{ }_\circ\left(-{1 \over 2}\tau\right) \hat{\cal
H} {\hat U}^\dagger_\circ\left(-{1 \over 2}\tau\right) e^{i\omega\tau} \ .
\end{equation}
We stress how the proposed time symmetrization gives rise to a fully symmetric Lindblad-like superoperator (expressed in terms of the operator $\hat{\cal L}$ only), compared to the strongly asymmetric Markov superoperator in (\ref{LvN-eff}).

If we now rewrite the new Markov superoperator in (\ref{LvN-eff-new2}) in our noninteracting basis $\lambda$, we obtain again the effective equation of motion in 
(\ref{LvN-eff-lambda}), but now the generalized scattering rates in (\ref{calP}) are replaced by the following symmetrized quantum scattering rates:
 \begin{equation}\label{calPtilde}
\tilde{\cal P}_{\lambda_1\lambda_2,\lambda'_1\lambda'_2} = {2\pi \over \hbar} H'_{\lambda_1\lambda'_1} H^{\prime *}_{\lambda_2\lambda'_2} \delta\left({\epsilon_{\lambda_1}+\epsilon_{\lambda_2}\over 2} - {\epsilon_{\lambda'_1}+\epsilon_{\lambda'_2}\over 2} \right) \ .
\end{equation}
The above scattering superoperator can be regarded as the quantum-mechanical generalization of the conventional Fermi's golden rule; indeed, in the semiclassical limit ($\lambda_1=\lambda_2,\lambda_1'=\lambda_2'$)
the standard formula in (\ref{P}) is readily recovered.

At this point a few comments are in order.
As discussed extensively in \cite{PRB}, also for the simplest case of a standard two-level system ---i.e., a generic quantum bit--- the standard Markov superoperator predicts a non-trivial coupling between level population and polarization described by the so-called $T_3$ contributions.
In contrast, for a two-level system coupled to its environment, the proposed quantum Fermi's golden rule does not predict any $T_3$ coupling term, thus providing a rigorous derivation of the well-known and successfully employed $T_1 T_2$ dephasing model \cite{SL}.
Moreover, it is imperative to stress that in the presence of a strong system-environment interaction the adiabatic decoupling investigated so far needs to be replaced by more realistic treatments, expressed via non-Markovian integro-differential equations of motion (i.e., with ``memory effects'') \cite{RMP}.
Again, while for purely atomic and/or photonic systems it is possible to identify effective non-Markovian evolution operators \cite{nonmarkovian}, for solid-state quantum devices this is still an open problem.

To summarize, we have critically reviewed the standard adiabatic or Markov procedure, showing its intrinsic failure in describing the proper quantum-mechanical evolution of a generic subsystem interacting with its environment. More specifically, we have shown that within the Markov approximation the density-matrix operator is not necessarily positive-definite, thus leading to highly non-physical results. To overcome this serious limitation, we have identified an alternative adiabatic procedure which (i) in the semiclassical limit reduces to the standard Fermi's golden rule, and (ii) describes a genuine Limblad evolution, thus providing a reliable/robust treatment of energy-dissipation and dephasing in state-of-the-art quantum devices.



\medskip\par\noindent

We are grateful to David Taj for stimulating and fruitful discussions.


\begin{thebibliography}{xx}
\bibitem{renormalizations}
Such approximation corresponds to neglect higher-order effects of the interaction Hamiltonian $\hat{\cal H}$ on the free system evolution ${\hat U}_\circ$. This is accomplished by neglecting contributions related to the commutator between the ``dressed'' evolution and the free one, i.e., 
$
\left[\left[{\hat U}^{ }_\circ, \hat{\cal H}\right],{\hat U}^{\dagger}_\circ\right] = 0
$.
Indeed, this is the condition needed in order to get the desired time-reversal symmetry
$
\hat{\cal H}^i(t) =
\hat{\cal H}^i(-t)
$.
\end{thebibliography}
\end{document}